\documentclass[letterpaper,12pt]{article}
\usepackage[utf8x]{inputenc}
\usepackage{amsmath,amsfonts,graphicx}
\usepackage{amssymb}
\usepackage{hyperref}
\usepackage[numbers,sort&compress]{natbib}
\usepackage[left=2.54cm,top=2.54cm,right=2.54cm,bottom=2.54cm,bindingoffset=0.0cm]{geometry}
\usepackage{authblk}
\usepackage{natbib}
\usepackage{setspace}
\usepackage{tabularx}

\usepackage{enumitem}
\setlist{nosep}

\begin{document}

\title{Multi-level computational methods for interdisciplinary research in the HathiTrust Digital Library}

\author[1,2]{Jaimie Murdock}
\author[1,3,+,*]{Colin Allen}
\author[1,2,4,5,+]{Katy B\"orner}
\author[2]{Robert Light}
\author[6]{Simon McAlister}
\author[10,+]{Andrew Ravenscroft}
\author[1,7]{Robert Rose}
\author[1]{Doori Rose}
\author[8]{Jun Otsuka}
\author[9,+]{David Bourget}
\author[10]{John Lawrence}
\author[10,+]{Chris Reed}
\affil[1]{\footnotesize Program in Cognitive Science, Indiana University, Bloomington, IN, USA}
\affil[2]{\footnotesize School of Informatics and Computing, Indiana University, Bloomington, IN, USA}
\affil[3]{\footnotesize Department of History \& Philosophy of Science \& Medicine, Indiana University, Bloomington, IN, USA}
\affil[4]{\footnotesize Indiana University Network Science Institute (IUNI), Bloomington, IN, USA}
\affil[5]{\footnotesize User-Centered Social Media, Department of Computer Science and Applied Cognitive Science, University of Duisburg-Essen, Duisburg, Germany}
\affil[6]{\footnotesize International Centre for Public Pedagogy (ICPuP), Cass School of Education \& Communities, University of East London, London, United Kingdom}
\affil[7]{\footnotesize Department of Mathematics, Indiana University, Bloomington, IN, USA}
\affil[8]{\footnotesize Department of Philosophy, Kyoto University, Kyoto, Japan}
\affil[9]{\footnotesize Department of Philosophy, University of Western Ontario, London, Ontario, Canada}
\affil[10]{\footnotesize Centre for g Technology, University of Dundee, Dundee, United Kingdom}
\affil[+]{These authors were the project leaders; see ``Author's Contributions'' for details of all contributions.}
\affil[*]{Corresponding author:  prof.colin.allen@gmail.com.  Address after August 2017: Department of History \& Philosophy of Science, University of Pittsburgh, 1017 Cathedral of Learning, 4200 Fifth Avenue, Pittsburgh, PA 15260, USA.}

\maketitle

\begin{abstract}
We show how faceted search using a combination of traditional classification systems and mixed-membership topic models can go beyond keyword search to inform resource discovery, hypothesis formulation, and argument extraction for interdisciplinary research. Our test domain is the history and philosophy of scientific work on animal mind and cognition. The methods can be generalized to other research areas and ultimately support a system for semi-automatic identification of argument structures. We provide a case study for the application of the methods to the problem of identifying and extracting arguments about anthropomorphism during a critical period in the development of comparative psychology. We show how a combination of classification systems and mixed-membership models trained over large digital libraries can inform resource discovery in this domain. Through a novel approach of ``drill-down'' topic modeling---simultaneously reducing both the size of the corpus and the unit of analysis---we are able to reduce a large collection of fulltext volumes to a much smaller set of pages within six focal volumes containing arguments of interest to historians and philosophers of comparative psychology. The volumes identified in this way did not appear among the first ten results of the keyword search in the HathiTrust digital library and the pages bear the kind of ``close reading'' needed to generate original interpretations that is the heart of scholarly work in the humanities. Zooming back out, we provide a way to place the books onto a map of science originally constructed from very different data and for different purposes. The multilevel approach advances understanding of the intellectual and societal contexts in which writings are interpreted.
\end{abstract}

\onehalfspacing
\section*{Introduction}

Just as Britain and America have been described as two nations separated by a common language, different academic disciplines often use the same words with divergent meanings~\cite{Demarest2015}. Interdisciplinary research thus poses unique challenges for information retrieval (IR). Word sense disambiguation~\cite{Mihalcea2010,Agirre2006}, differing publication practices across disciplines~\cite{Borgman2002,Cronin2003,Holmberg2014} and disjoint authorship networks~\cite{Karki1996} pose special challenges to information retrieval for interdisciplinary work. When the dimension of time is added, terminological shifts~\cite{Hamilton2016,Hamilton2016a}, changing citation standards~\cite{Kaplan1965,Liu1993,Lariviere2008,DeRijcke2016}, and shifting modes of scholarly communication~\cite{Odlyzko2002,Borgman2002,Cronin2003,Evans2008} all amplify the challenges for IR to serve the need of interdisciplinary scholars. 

Widespread digitization of monographs and journals by HathiTrust~\cite{York2009,Christenson2011} and Google Books~\cite{Coyle2006,Vincent2007} enable new longitudinal studies of change in language and discourse~\cite{Lariviere2008,Altmann2009,Michel2011,Cocho2015,Hamilton2016,Hamilton2016a}, an approach known as ``distant reading''~\cite{Moretti2013}. These data-driven distant readings contrast with ``close readings'', in which short passages and particular details are emphasized for scholarly interpretation. Newly digitized materials, which enable distant reading, differ from born-digital scholarly editions in three key ways: First, the reliance on optical character recognition (OCR) over scanned page images introduces noise into the plain-text representations of the text. Second, the unstructured text does not contain any markup that may differentiate page header and footer information, section headings, or bibliographic information from the main text.  Finally, metadata is often automatically extracted and lacks the provenance information important to many humanities scholars. Researchers seeking to marry these ``distant readings'' to more traditional ``close readings'' are impacted by these factors~\cite{Underwood2014a}.

Our goal is to develop computational methods for scholarly analysis of large-scale digital collections that are robust across both the technological inconsistency of the digitized materials and the variations of meaning and practice among fields and across time. A further goal of our approach is that these methods should inform interdisciplinary research by suggesting novel interpretations and hypotheses. The methods should support scholars who wish to drill down from high level overviews of the available materials to specific pages and sentences that are relevant for understanding the various responses of scholars and scientists to contentious issues within their fields. 

In this paper, we provide a case study that focuses on meeting these challenges within the interdisciplinary field of History and Philosophy of Science (HPS). HPS must not only bridge the humanities and the sciences, but also the temporal divide between historically-significant materials and the present~\cite{Kuhn1979,Laudan1986,Hacking1990,Weingart2015}. We show how faceted search using a combination of traditional classification systems and mixed-membership models can go beyond keyword search to inform resource discovery, hypothesis formulation, and argument extraction in our test domain, delivering methods that can be generalized to other domains.  

Using a novel approach of drill-down topic modeling---simultaneously reducing both the size of the corpus and the unit of analysis---we demonstrate how a set of 1,315 fulltext volumes obtained by a keyword search from the HathiTrust digital library is progressively reduced to six focal volumes that did not appear in the top ten results in the initial HathiTrust search. Topic modeling of these volumes at various levels, from whole book down to individual sentences, provides the contexts for word-sense disambiguation, is relatively robust in the face of OCR errors, and ultimately supports a system for semi-automatic identification of argument structure.  We show how visualizations designed for macroanalysis of disciplinary scientific journals can be extended to highlight interdisciplinarity in arguments from book data~\cite{Borner2012}. This guides researchers to passages important for the kind of ``close reading'' that lies at the heart of scholarly work in the humanities, supporting and augmenting the interpretative work that helps us understand the intellectual and societal contexts in which scientific writings are produced and received.

While the extension of computational methods such as these to various questions in the humanities may eventually provide ways to test specific hypotheses, the main focus of such research is likely to remain exploratory and interpretative, in keeping with the humanities themselves~\cite{Underwood2014a,Rockwell2016}. This approach nevertheless shares something with the sciences: it is experimental to the extent that it opens up a space of investigation within which quantitatively defined parameters can be systematically varied and results compared. Such exploratory experimentation is common not just in the social sciences, but also in the natural sciences~\cite{Steinle1997,Waters2007}. 

Our study consisted of six stages. (1) We used a keyword search of the HathiTrust collection to generate an initial corpus and we used  \emph{probabilistic topic models} on these volumes. (2) We exploited the \emph{mixed-membership} property of the topic models to identify the multiple contexts of the selected volumes and reduce the original search space even further. (3) Because topic models define the notion of a document flexibly, we drilled down further by constructing \emph{page-level topic models} of the reduced set of volumes selected at the previous stage. (4) We used the page-level results to rank books and select pages from them for closer analysis, demonstrating an approach to semi-automatic \emph{argument extraction} which showcases the interpretive results of our search process. (5) We exploited the close reading of arguments for exploratory investigation of drilling down even further, to \emph{sentence-level topic modeling} within a single volume. (6) We used \emph{scientific mapping} to locate relevant volumes~\cite{Borner2010}. Because current science maps represent journal data, and data overlays are created based on journal names, we needed to construct a \emph{classification crosswalk} from the UCSD Map of Science to the Library of Congress Classifications of these journals, finally allowing us to project books onto the science map. 

We assessed success in our case study in three ways: (1) by the effectiveness of the process in leading non-experts to drill down to highly-relevant content in a very large collection of books; (2) by the ability of this process to spotlight a somewhat forgotten woman scientist who is important to the history of psychology; (3) by the capacity of the process to lead domain experts to a surprising discovery about the breadth of species discussed in these historical materials, thus enriching the historical context for current discussions of intelligence in microscopic organisms~\cite{Nakagaki2000,Reid2016}. Our assessments are qualititative rather than quantitative in nature, but they are appropriate given current limitations in quantitative assessments of the quality of topic models~\cite{Chang2009,Blei2012,Lee2017}.

\subsection*{Related Work}
%TODO: CA: Review new paragraphs below on topic modeling and related work
The use of topic models for information retrieval is not itself novel, having prior general applications \cite{Wei2006,Medlar2017}, scientific applications \cite{Griffiths2004,Hall2008}, and humanities applications \cite{Tangherlini2013}. Similar to our approach, some of these applications support finer-grained retrieval by remodeling a subset of the corpus.  The key novelty of our approach, is that we simultaneously alter the granularity of the documents in our models as we go from modeling books in collections, to pages in books, to sentences in pages. 

Previous studies indicate a general consensus that human judgments about what makes a ``good'' topic are generally convergent. However, human judgment does not typically correlate well with quantitative measures of model fit \cite{Chang2009}, suggesting that people are interpreting the topics using as-yet poorly understood semantic criteria. Furthermore, variation among people in their interpretation of topic quality may be dependent upon expertise. Some topics that are poorly-rated by non-experts may in fact be judged highly coherent by experts who understand why certain documents have high membership in the topic, in contrast to non-experts who focus solely on the highest-probability terms in the topic without knowledge of the underlying corpus \cite{Lee2017}. Interactive topic modeling \cite{Hu2014} approaches this issue by introducing human-in-the-loop topic selection and biasing measures that increase human judgment of topic model fitness. Our drill-down topic modeling approach does not require human feedback during the modeling stage, but during the corpus selection phase. This reduces the training cost of our approach and makes it more accessible for exploratory search. 

The use of visualization techniques in information retrieval is well documented: Doyle's ``Semantic Road Maps for Literature Searchers'' explicitly justified the use of visualization as a summary of scientific literature, in particular as a time-saving measure by quickly showing relevant features of a document \cite{Doyle1961}. Doyle also emphasizes that even if a visualization is itself static, it is the result of a dynamic process of iterative remodeling and learning from new data. The UCSD Map of Science is a basemap that needs to be learned---just like a geographic map of the world---but that can subsequently be used to quickly gain an overview of the topical distribution of documents\cite{Borner2012}. Visualization of semantic models is also well-documented, especially for topic models \cite{Chaney2012,Chuang2015,Murdock2015}. Prior models, including the results of LSA, word co-occurrence, and other semantic analyses were also visualized (see \cite{Borner2010} for a timeline). The last step of the workflow described in this paper uniquely projects a topic model analysis onto a visualization base layer derived from different data (journal citation links) for different purposes (visualizing the citation structure of current science). While the sorted lists we provide below are useful for determining what to read next, visualization helps users to understand patterns, trends, and outliers, supporting quick evaluation of which items are most relevant to their interests.

\section*{Materials}
\urlstyle{same}

\subsection*{HathiTrust Digital Library}
The HathiTrust Digital Library is a collaboration between over ninety institutions to provide common access and copyright management to books digitized through a combination of Google, Internet Archive, and local initiatives. As of October 24, 2016, it consisted of over 14.7 million volumes represented both as raw page images and OCR-processed text\footnote{\url{https://www.hathitrust.org/statistics_info}}. 

Due to copyright concerns, fulltext access to page images and their OCR-processed counterparts is given only to pre-1928 materials, which are assumed to be in the public domain in the United States.\footnote{During the funding period, even summary data describing the fulltext of post-1928 materials were impossible to access for computational analysis from the HathiTrust. Recently, the HathiTrust Research Center (HTRC) Data Capsule has been developed to enable tightly restricted access to features extracted from in-copyright materials~\cite{Zeng2014}.} When the work described in this paper was initiated in 2012, the public domain portion of the HathiTrust consisted of approximately 300,000 volumes. At the end of the funding period in 2014, the public domain consisted of 2.1 million volumes. As of October 24, 2016, that number stood at 5.7 million volumes, and it has continued to grow since then. 

While the corpus size has increased more than 20-fold, the methods presented in this paper are aimed to reduce the portion of the corpus for analysis. For example, the first step described below involves topic modeling the results of a \emph{keyword search}, resulting in a corpus of 1,315 volumes (which we referred to as $HT1315$). Using the same query on October 24, 2016, we returned 3,497 volumes. Both of these datasets are computationally-tractable for topic modeling on modern workstations, in contrast (for example) to the 1.2 terabyte HTRC Extracted Features Dataset, derived from 4.8 million volumes~\cite{Capitanu2015}. The methods described in detail below further reduced the $HT1315$ corpus to a smaller corpus of 86 volumes ($HT86$) which we modeled at the page level. This corpus was then further analyzed and refined to a 6-volume collection for argument mapping ($HT6$).

\subsection*{Stop Lists}
Before analyzing the texts, it is common to apply a `stop list' to the results, which excludes words that are poor index terms \cite{Fox1989}. Frequently, these are high-frequency words such as articles (`a', `an', `the'), prepositions (`by', `of', `on'), and pronouns (`he', `she', `him'), which contain little predictive power for statistical analysis of semantic content \cite{Luhn1957}. We use the English language stop list in the Natural Language Toolkit, which contains 153 words \cite{nltk}. Additionally, we filtered words occurring five or fewer times, which both excludes uncommon words and infrequent non-words generated by OCR errors. We also developed custom methods for stripping headers and footers from the OCRed pages provided by the HathiTrust, cleaning up hyphenated words crossing lines and page breaks, and obtaining volume metadata.\footnote{Source code available at https://github.com/inpho/vsm/blob/master/vsm/extensions/htrc.py.}

\subsection*{UCSD Map of Science}
For our macroanalysis, we want to see how our selected texts divide among the different academic disciplines. As a base map for the disciplinary space (analogous to a world map for geospatial space), we use the UCSD Map of Science \cite{Borner2012} which was created by mining scientific and humanities journals indexed by Thomson Reuters' Web of Science and Elsevier's Scopus. The map represents 554 sub-disciplines---e.g., Contemporary Philosophy, Zoology, Earthquake Engineering---that are further aggregated into 13 core disciplines, appearing similar to continents on the map---e.g., Biology, Earth Sciences, Humanities. Each of the 554 sub-disciplines has a set of journals and keywords associated with it.

\subsection*{Library of Congress Classification Outline (LCCO)}
The Library of Congress Classification Outline (LCCO) is a system for classifying books, journals, and other media in physical and digital libraries. It is different from the Library of Congress Control Number (LCCN), which provides an authority record for each volume. The HathiTrust stores the LCCN, which we then use to query the Library of Congress database for the call number, which contains the LCCO, providing us with a disciplinary classification for each volume in the $HT1315$, $HT86$, and $HT6$ datasets.

\subsection*{Target Domain: History and Philosophy of Scientific Work on Animal Cognition}
Our specific test domain is the history and philosophy of scientific work on animal cognition~\cite{Allen1999,sep-cognition-animal,sep-consciousness-animal}. We aimed to identify and extract arguments about anthropomorphism from a relevant subset of the scientific works published in the late 19th and early 20th century. This period represents a critical time for the development of comparative psychology, framed at one end by the work of Charles Darwin and at the other end by the rise of the behaviorist school of psychology (see \cite{Boakes1984} for a full historical review). Using the methods described in this paper, we progressively narrowed the 300,000 volumes to a subset of 1,315 selected for topic modeling at the full-volume level, then 86 of these selected for page-level topic modeling, and then 6 specific volumes selected for manual analysis of the arguments.

The term `anthropomorphism' itself illustrates the problem of word sense disambiguation. In theological and anthropological contexts, `anthropomorphism' refers to the attribution of human-like qualities to gods. In the animal cognition context, it refers to the projection of human psychological properties to animals. Given the theological controversy evoked by Darwin, our inquiry demands our system be robust in partitioning these separate discourses.

\section*{Methods}
\subsection*{Methods Overview}
\begin{figure}
\includegraphics[width=\textwidth]{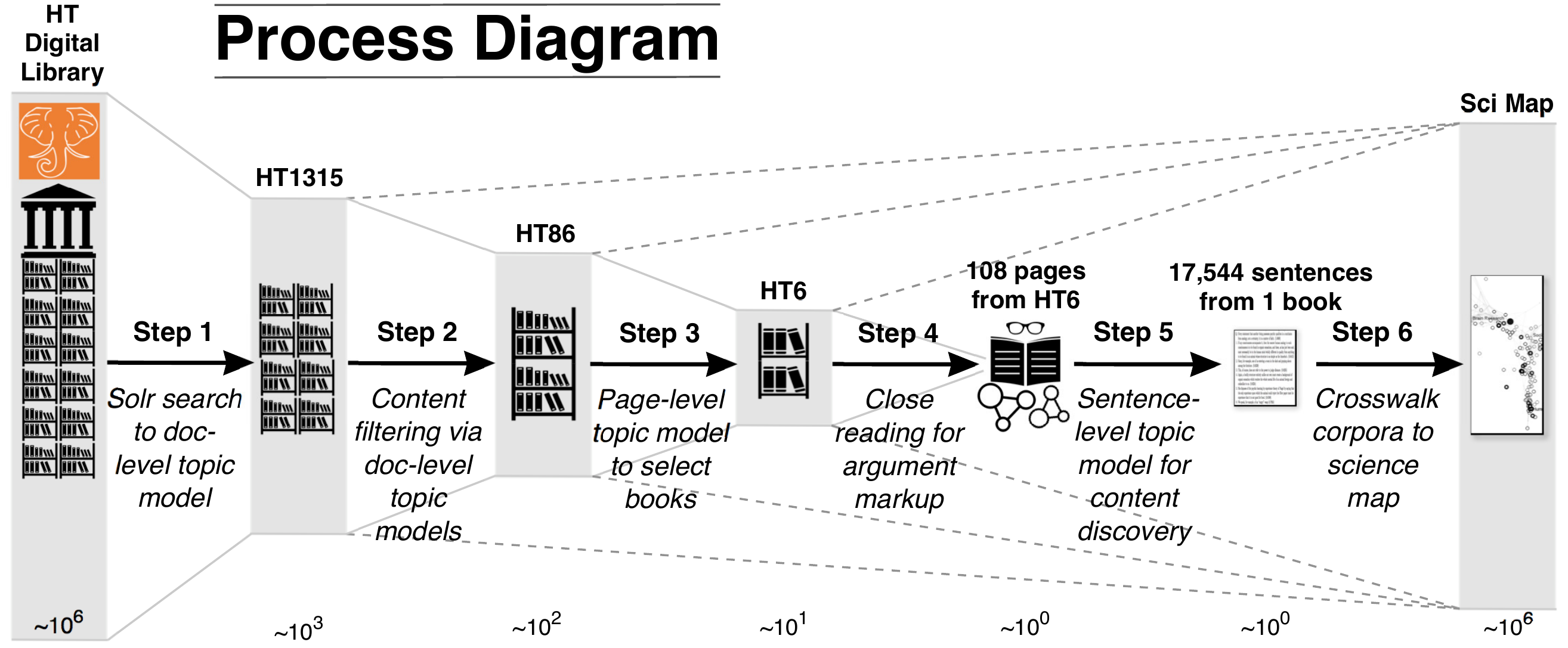}
\caption{Schematic rendering of the six-step process that sequentially drills down from macroscopic ``distant reading'' to microscopic ``close reading'' before zooming back out to the macroscopic scale at the final step. The approximate orders of magnitude of the datasets either side of each processing step are shown below the icons as powers of 10 of book/fulltext-sized units, and grey bars representing the data are scaled logarithmically.}
\label{fig:process}
\end{figure}

We followed a six-stage process, summarized in Fig.~\ref{fig:process}. Each step is described in more detail further below. We introduce them briefly here:
\begin{enumerate}
\item LDA Topic modeling of a subset of volumes from the HathiTrust Digital Library selected by a keyword search, treating each volume as the unit document for the LDA process.
\item Querying the model to further reduce the original set of documents to a more relevant subset for our HPS objectives. 
\item Drill-down LDA topic modeling on the smaller set treating individual pages as the unit documents, using this page-level model to select pages for further analysis.
\item Mapping of arguments on the selected pages by manual analysis, supported by the enhanced Online Visualisation of Arguments (OVA+) tool \cite{Lawrence2012}.
\item LDA topic modeling of single books, treating each \emph{sentence} as document unit.
\item Mapping identified volumes onto UCSD Map of Science via a crosswalk from Library of Congress classification data to the journals used to construct the basemap. 
\end{enumerate}

%\textbf{Add more technical details here -- pick a document, pick a topic, pick a word - generative vs. predictive, etc..} % TODO: Add more technical details here

\subsection*{Detailed Methods}
\subsubsection*{1. From Keyword Search to Probabilistic Topic Modeling}

We reduced the number of volumes to be routed to the topic modeling process by conducting a keyword search in the HathiTrust  collection using the HathiTrust's Solr index. We searched using terms intended to reduce the hundreds of thousands of public domain works to a set of potentially relevant texts that could be efficiently modeled with the available computing resources. Specifically, we searched for ``Darwin'', ``comparative psychology'', ``anthropomorphism'', and ``parsimony''. While the specificity of our query may be seen as too restrictive, we emphasize (a) that we are following an exploratory research paradigm - we are not narrowing in on a particular fact, but rather surveying the available literature at the intersection of our interest in the history and philosophy of animal mind and cognition, and (b) the results of the keyword search were not specific enough to make the topic modeling redundant. Because we retrieved 1,315 volumes from the HathiTrust by this method, we refer to this corpus as $HT1315$. (More details can be found in the Results section below.)

\emph{Probabilistic topic models}~\cite{Blei2012} are a family of mixed-membership models that describe documents as a distribution of topics, where each topic is itself a distribution over all words in a corpus. Topic models are \emph{generative} models, that we interpret as providing a theory about context blending during the writing process \cite{Murdock2015darwin}. 

Corpus preparation begins by treating each document as a bag of words. Common function words (prepositions, conjunctions, etc.) were filtered out using the NLTK stopword list for English, and rare words were filtered using a lower bound of 5 occurrences in the corpus. To construct the topic models used in this study, we use \emph{Latent Dirichlet Allocation} (LDA---\cite{Blei2003}) with priors estimated via Gibbs sampling \cite{Griffiths2004} as implemented in the InPhO Topic Explorer \cite{Murdock2015}. 

The topic-modeling process begins by assigning random probabilities to the word-topic distributions and to each of the topic-document distributions. These prior distributions are then jointly sampled to generate estimates of the likelihood of observing the actual documents. These estimates are used to adjust the prior distributions in accordance with Bayes' rule. We ran this generate-and-test procedure iteratively for 1000 cycles, a number of interations at which the distributions become relatively stable.  Hyperparameters $\alpha$ and $\beta$ control the word-topic and topic-distributions. We set them equal to $0.1$, representing the expectation that each document should be weighted toward a mixture in which a relatively small subset of the available topics ($k$) dominate, and that topics should similarly be dominated by a relatively small proportion of the available words in the corpus. We initially modeled the $HT1315$ volumes using four different values for $k$, i.e., $k \in \{20,40,60,80\}$. 

\subsubsection*{2. Querying the models}
At the end of the modeling process, each document is represented as a probability distribution over the $k$ topics. We manually inspected the topics generated for the different values of $k$ and determined that while all four of the models produced interpretable results, $k=60$ provided the best balance between specificity and generality for our HPS goals.

We use the topic model to further narrow the search by querying topics with a combination of words. We do this by finding the topic or topics with the highest sum of the probabilities for each word. By a combination of trial and error, we found that a topic query combining ‘anthropomorphism’, ‘animal’, and ‘psychology’ produced more relevant topics and any term alone.

Using three topics identified in this way, we filtered the originally modeled set of books to a much smaller sub-corpus.  The topic-document and  word-topic distributions can be treated as vectors in their respective topic and word spaces. Thus it is possible to take the widely-used measure of vector cosines to assess similarity between topics and volumes. We computed the cosine distance between each of the three topics and the book's mixture of topics represented in the model.  We summed these three distances and filtered them at the threshold of 1.25, yielding a smaller 86-book corpus ($HT86$) for more detailed analysis. The cutoff was chosen by trial and error, manually inspecting the titles of the first few books excluded at a given threshold.

%Subsequently, we used all three topics (10, 16, and 26) to filter relevant books from the original set of 1,315 books.

\subsubsection*{3. Drill down to page level}

The notion of a ``document'' in LDA topic modeling is flexible. One can consider a full volume as a single document with a particular topic distribution. However, finer-grained models can also be made, in which each page, paragraph, or sentence receives its own topic distribution. Since OCR document scans in the HathiTrust have very little structural information---there is no encoding for section headings or paragraph breaks, let alone chapter breaks---the printed page was the next level below the full volume that we could reliably recover.

Hence, we re-modeled the $HT86$ set at the level of individual pages again using LDA topic modeling for values of $k \in \{20,40,60,80\}$, parameterized as before, towards the goal of identifying arguments in text by ``zooming in'' to select books which had a high number of apparently relevant pages. For the sake of direct comparison to results reported above with the $HT1315$ model, we probed the $k=60$ page-level model with `anthropomorphism' as the query term alone, and in combination with other terms 'animal' and 'psychology' used previously. This identified one topic as most relevant to our project (see Results for details). We ranked volumes from the $HT86$ corpus according to which had the most pages among the top 800 highest ranked pages according to this topic and selected the top six volumes for the next step of the process ($HT6$). (The choice of six here was limited by time and resources allocated to the manual extraction of arguments detailed in the next section.) 

\subsubsection*{4. Argument Extraction: From Pages to Arguments}

 The selected pages were annotated using the Argument Interchange Format ontology (AIF \cite{Chesnevar2006}), which defines a vocabulary for describing arguments and argument networks. One of the coauthors [SM], who is not a domain expert, identified arguments using a semi-formal discourse analysis approach (informed by~\cite{ravenscroft2000, pilkington2016}), and following a rubric established by the project PIs with HPS expertise [CA, DB, and AR]. The rubric supported identification of arguments based on their content and propositional structure, where this was also aided by noting argument signifiers in the texts, such as `because', `hence', `therefore', etc.\footnote{Additional details about the rubric can be found in section 2.3.3 of \cite{McAlister2014}.} This allowed us then to generate argument maps in the form of AIF annotated documents constructed with OVA+\footnote{The enhanced Online Visualization of Arguments tool OVA+ is available at \url{http://ova.arg-tech.org/}; see also \cite{Lawrence2012}}, an application which links blocks of text using argument nodes. OVA+ provides a drag-and-drop interface for analyzing textual arguments. It also natively handles AIF structures. Each argument was divided into propositions and marked up as a set of text blocks. These text blocks containing propositions were linked to propositions that they support, or undercut, to create argument maps.  OVA+ thus produces a visual representation of the structure of each argument.

\subsubsection*{5. Drilling Down Again: From Arguments to Sentences}

To further investigate the utility of combining distant reading methods with close reading, we applied topic modeling to the sentences within a single volume. For this test we selected Margaret Washburn's \emph{The Animal Mind} textbook~\cite{Washburn1908} because it was top-ranked for topical content in $HT6$. We applied LDA topic modeling to its 17,544 sentences, treating this set of sentences as a collection of documents. To explore the power of topic modeling to identify latent but meaningful relationships at the micro-level, we arbitrarily chose a sentence from an Argument extracted from the Washburn set and used it to query the sentence-level model of \emph{The Animal Mind} for the most similar sentences using the cosine of the sentence-topic vectors.
%Argument 15%

\subsubsection*{6. Zooming Out: Macroanalysis by Science Mapping}

At the final step, we created a visualization of the retrieved books overlaid on the UCSD Map of Science~\cite{Borner2010}, to help understand the distribution of the retrieved books with respect to scientific disciplines. 

In previous research, new datasets have been overlaid on the UCSD map by matching records via journal names or keywords to the 554 sub-disciplines. However, our present study is the first time that book data have been overlaid on a science map. To accomplish this, we constructed a \emph{classification crosswalk} to align the journal-based sub-disciplines with a book classification system. The Library of Congress Classification Outline (LCCO) provides a hierarchical disciplinary taxonomy similar to that of the UCSD Map of Science. By using the Library of Congress Control Numbers (LCCN) assigned to each of the 25,258 journal sources in the UCSD Map of Science, we were able to use the hierarchical structure of the LCCO to assign a likelihood to any given book LCCN belonging to a particular USCD sub-discipline.

A number of items in the HathiTrust collection never received LCCNs. For example, university library collections frequently contain course bulletins that are not catalogued by the Library of Congress. We removed the uncatalogued items and projected the remaining volumes onto the UCSD map of science. We assigned each remaining book in $HT1315$ a UCSD sub-discipline based on its LCCN.

\section*{Results: A Case Study}

In this section we describe the application of these methods to a case study in the History \& Philosophy of Science (HPS), specifically in the history of comparative psychology. When we began the study the HathiTrust digital library provided access to the full texts of just over 300,000 public domain works. The keyword-based search for items of interest reduced this set to a corpus of 1,315 volumes published between 1800 and 1962,\footnote{Publication dates after 1928 correspond to items in the public domain, such as government reports and university course bulletins.} our $HT1315$ corpus.\footnote{A list of titles and HT handles is provided in the supplemental materials. Because the HT collection has changed over time, this exact set of results cannot be recreated by doing the same keyword search at hathitrust.org (see \url{http://bit.ly/1LBbqnS}). Currently there are over 5.5 million public domain works in the collection (see \url{https://www.hathitrust.org/visualizations_dates_pd}). The same query conducted in August 2015 yielded 3,027 full-text results.}

Table~\ref{table:ht1315-anthro-topics} shows the top topics when the  $k=60$ topic model is queried using the single word `anthropomorphism'. The topic model checking problem \cite{Blei2012}---i.e., how to assess the quality of the model’s topics---remains an important open problem in topic modeling. Nevertheless, most of the topics in the model can be quickly summarized. Inspection of this list indicates that `anthropomorphism' relates most strongly to a theological topic (38), a biological topic (16), a philosophical topic (51), an anthropological topic (58), and a child development topic (12). The topic model thus serves to disambiguate the different senses of 'anthropomorphism', especially between contexts where the discussion is about anthropomorphized deities (38) and contexts where it is about nonhuman animals (16), with the second topic being the most obvious attractor for researchers interested in comparative psychology. The second-to-last topic (1) is targeted on bibliographic citations, and is dominated by bibliographic abbreviations and some common German and French words that were not in the English language stop list used during initial corpus preparation. Although from one perspective this may seem like a `junk' topic, this topic is nonetheless very useful to a scholar seeking citations buried in the unstructured pages in the corpus.

\begin{table}
%\begin{center}
% \textbf{Topics Ranked by Similarity to `anthropomorphism'}
\begin{tabularx}{\textwidth}{|c|X|}
\hline
\textbf{Topic}	& \textbf{10 most probable words from topic} \\ \hline
38 &	god, religion, life, man, religious, spirit, world, nature, spiritual, divine \\
\textbf{16} &	\textbf{animals, evolution, life, animal, development, man, species, cells, living, theory} \\
51 &	philosophy, nature, knowledge, world, thought, idea, things, reason, truth, science \\
58 &	man, among, tribes, primitive, men, people, also, races, women, race \\
12 &	child, children, first, development, movements, play, life, little, mental, mother \\
21 &	social, life, new, mind, upon, individual, human, mental, world, subfield \\
11 &	motion, force, must, forces, matter, changes, us, parts, like, evolution \\
1 &	pp, der, vol, die, de, des, und, ibid, university, la \\
31 &	gods, religion, p, name, see, god, india, ancient, one, worship \\ 
\hline
\end{tabularx}
%\end{center}
\caption{Topics ranked by similarity to `anthropomorphism' in the $HT1315$ corpus. Topic 16 (highlighted with bold text) is highly relevant to the inquiry.}
\label{table:ht1315-anthro-topics}
\end{table}

\begin{table}
%\begin{center}
% \textbf{Topics Ranked by Joint Similarity to `anthropomorphism', `animal', and `psychology'.}
\begin{tabularx}{\textwidth}{|c|X|}
\hline
\textbf{Topic}	& \textbf{10 most probable words from topic} \\ \hline
\textbf{26} &	\textbf{consciousness, experience, p, psychology, process, individual, object, activity, relation, feeling} \\
\textbf{16} &	\textbf{animals, evolution, life, animal, development, man, species, cells, living, theory} \\
\textbf{10} &	\textbf{animals, water, animal, food, birds, one, leaves, insects, species, many} \\
47 &	college, university, professor, school, law, work, students, degree, education, new \\
49 &	subfield, code, datafield, tag, ind2, ind1, b, d, c, controlfield \\
1 &	pp, der, vol, die, de, des, und, ibid, university, la \\
12 &	child, children, first, development, movements, play, life, little, mental, mother \\
58 &	man, among, tribes, primitive, men, people, also, races, women, race \\
21 &	social, life, new, mind, upon, individual, human, mental, world, subfield \\
2 &	test, tests, age, group, children, mental, table, per, cent, number \\
\hline
\end{tabularx}
%\end{center}

\caption{Topics ranked by similarity to `anthropomorphism', `animal', and `psychology' in the $HT1315$ corpus. Topics 26, 16, and 10 (highlighted with bold text) were used to derive the $HT86$ corpus, as they were most relevant to the inquiry.}
\label{table:ht1315-3words-topics}
\end{table}
Table~\ref{table:ht1315-3words-topics} shows the top topics returned by querying the $k=60$ model of $HT1315$ using `anthropomorphism', `animal', and `psychology' to construct the query. This new query reveals two relevant topics (numbers 26 and 10) that were not returned using `anthropomorphism' alone. The top ten documents found by querying the model using these two topics in combination with the previously noted topic 16 is shown in Table~\ref{table:ht86-docs}. By selecting from the continuation of this list up to a threshold of 1.25 on the aggregated distance measure, we reduced the number of volumes of interest from 1,315 to 86, constituting the $HT86$ corpus.

\begin{table}[ht]
\begin{center}
\end{center}

\begin{tabularx}{\textwidth}{|Xr|}
\hline
\textbf{Document} & \textbf{Distance} \\ \hline
Secrets of animal life & 0.87689\\
Comparative studies in the psychology of ants and of higher \ldots & 0.88814\\
The colours of animals, their meaning and use, especially \ldots & 0.98445\\
The foundations of normal and abnormal psychology &	0.99833\\
The bird rookeries of the Tortugas & 1.00286\\
Mind in animals & 1.00294\\
Ants and some other insects; an inquiry into the psychic \ldots & 1.00504\\
Systematic science teaching: a manual of inductive \ldots & 1.01040\\
The riddle of the universe at the close of the 19th C. & 1.01450\\
The colour-sense: its origin and development. & 1.02795\\
\hline
\end{tabularx}

\caption{Book titles ranked by proximity of the full texts to topics 10, 16, and 26 in the $k=60$ model of the $HT1315$ corpus.}
\label{table:ht86-docs}
\end{table}

%uc2.ark:/13960/t6057f659
%uc2.ark:/13960/t7wm15g73
%wu.89095158218
%uc2.ark:/13960/t9t14w82w
%mdp.39015009245195
%loc.ark:/13960/t9m33nm99
%uc2.ark:/13960/t5w66bs1h
%uc2.ark:/13960/t3bz6393c
%uc2.ark:/13960/t3125rr5w
%uc2.ark:/13960/t7gq6st77

\begin{table}
%\textbf{Topics Ranked by Similarity to `anthropomorphism'}
\begin{tabularx}{\textwidth}{|c|X|}
\hline
\textbf{Topic}	& \textbf{Top Ten Most Probable Words from Topic} \\ \hline
18 &	god, religion, evolution, religious, man, human, science, world, christian, belief \\
3 &	mind, man, facts, life, evolution, instinct, subjective, instincts, organic, development \\
1 &	animal, animals, may, stimulus, experience, would, instinct, reaction, one, stimuli \\
51 &	sense, sensation, qualities, touch, perception, sensations, extension, sight, senses, us \\
\hline
\end{tabularx}
\caption{Topics ranked by similarity to `anthropomorphism' in the $HT86$ corpus, as modeled at the page level.}
\label{table:ht86-anthro-topics}
\end{table}

The result of querying the $k=60$ page-level model of the $HT86$ corpus with the single query word `anthropomorphism' is shown in Table~\ref{table:ht86-anthro-topics}. (Topic numbers are arbitrary and do not correlate across the $HT86$ and $HT1315$ models.) Although a theological topic (18) is again at the top of the list, it is clear that biological and psychological topics have become more prevalent in the $HT86$ model. Even within Topic 18, `evolution' and `science' are now among the ten highest probability words indicating that the topic is closer to a ``religion and science'' topic than the more general religion Topic 38 from the $HT1315$ model (Table~\ref{table:ht1315-anthro-topics}), and reflecting the narrower range of books in the $HT86$ subset.

\begin{table}

%\textbf{Topics Ranked by Similarity to `anthropomorphism', `animal', and `psychology'}
\begin{tabularx}{\textwidth}{|c|X|}
\hline
\textbf{Topic}	& \textbf{Top Ten Most Probable Words from Topic} \\ \hline
1 &	animal, animals, may, stimulus, experience, would, instinct, reaction, one, stimuli \\
51 &	sense, sensation, qualities, touch, perception, sensations, extension, sight, senses, us \\
18 &	god, religion, evolution, religious, man, human, science, world, christian, belief \\
3 &	mind, man, facts, life, evolution, instinct, subjective, instincts, organic, development \\ \hline 
\end{tabularx}

\caption{Topics ranked by similarity to `anthropomorphism', `animal', and `psychology' in the $HT86$ corpus.}
\label{table:ht86-3words-topics}
\end{table}

Using `anthropomorphism', `animal' and `psychology' in combination to query the $k=60$ $HT86$ model, topic 1 is the highest ranked topic (Table~\ref{table:ht86-3words-topics}). In comparison to the earlier topics 10 and 16 from the $HT1315$ results in Table~\ref{table:ht1315-3words-topics}, this topic has more terms relevant to psychology (i.e., stimulus, experience, instinct, reaction), suggesting that for the purposes of locating specific pages in $HT86$ collection that are relevant to our HPS interests, topic 1 provides the best starting point. Table \ref{table:page-similarity} shows the first rows of a list of 800 highest ranked pages from $HT86$ using topic 1 as the query.

\begin{center}
\begin{table}

\begin{tabularx}{\textwidth}{|X|c|}
\hline
\textbf{Document} & \textbf{Distance} \\ \hline
The animal mind, 1st ed., p. 43 & 0.04414 \\
The animal mind, 2nd ed., p. 47	& 0.04552 \\
The animal mind, 2nd ed., p. 263 & 0.10360 \\
The animal mind, 2nd ed., p. 16	& 0.12336 \\
The animal mind, 2nd ed., p. 71 & 0.15828 \\
The animal mind, 1st ed., p. 219 & 0.16288 \\
The animal mind, 1st ed., p. 232 & 0.16674 \\
The animal mind, 1st ed., p. 57	& 0.18380 \\
The animal mind, 1st ed., p. 72 & 0.22610 \\
Mind in the lower animals, p. 179 & 0.23408 \\
\hline 
\end{tabularx}

\caption{Pages ranked by similarity to Topic 1.}
\label{table:page-similarity}
\end{table}
\end{center}

None of the six volumes from $HT86$ collection which had the most pages in the top 800 highest-ranked pages had  appeared among the top 10 keyword search results in the original Solr search in the HathiTrust collection. These volumes formed the $HT6$ collection:
\begin{enumerate}
\item \emph{The Animal Mind: A Textbook of Comparative Psychology}, 1908 (first edition), by Margaret Floy Washburn, psychologist. Washburn's textbook was foundational for comparative psychology and she is notable as the second woman to be president of the American Psychological Association.
\item \emph{Comparative studies in the psychology of ants and of higher animals}, 1905, a monograph by Erich Wasmann, an entomologist who only partly accepted evolution within species, rejecting common descent, speciation via natural selection, and human evolution.
\item \emph{The Principles of Heredity}, 1906, a scientific monograph by G. Archdall Reid, a physician who argued against the Lamarckian idea of inheritance of acquired characteristics. 
\item \emph{General Biology}, 1910, a text book by James G. Needham, entomologist and limnologist.
\item \emph{The Nature and Development of Animal Intelligence}, 1888, a compilation of articles by Wesley Mills, physiologist, physician and veterinarian.
\item \emph{Progress of Science in the Century}, 1908, a book on the history of science for general readers by J. Arthur Thomson, naturalist.
\end{enumerate}

These books were written by two Americans (Washburn and Needham), two Scots (Reid and Thomson), a Canadian (Mills), and an Austrian (Wasmann). They provide a broad array of perspectives on animal intelligence and psychology, from specialist monographs to textbooks to general-audience nonfiction, spanning both pro-Darwinian and anti-Darwinian viewpoints. 

\begin{figure}
\includegraphics[width=7.0in]{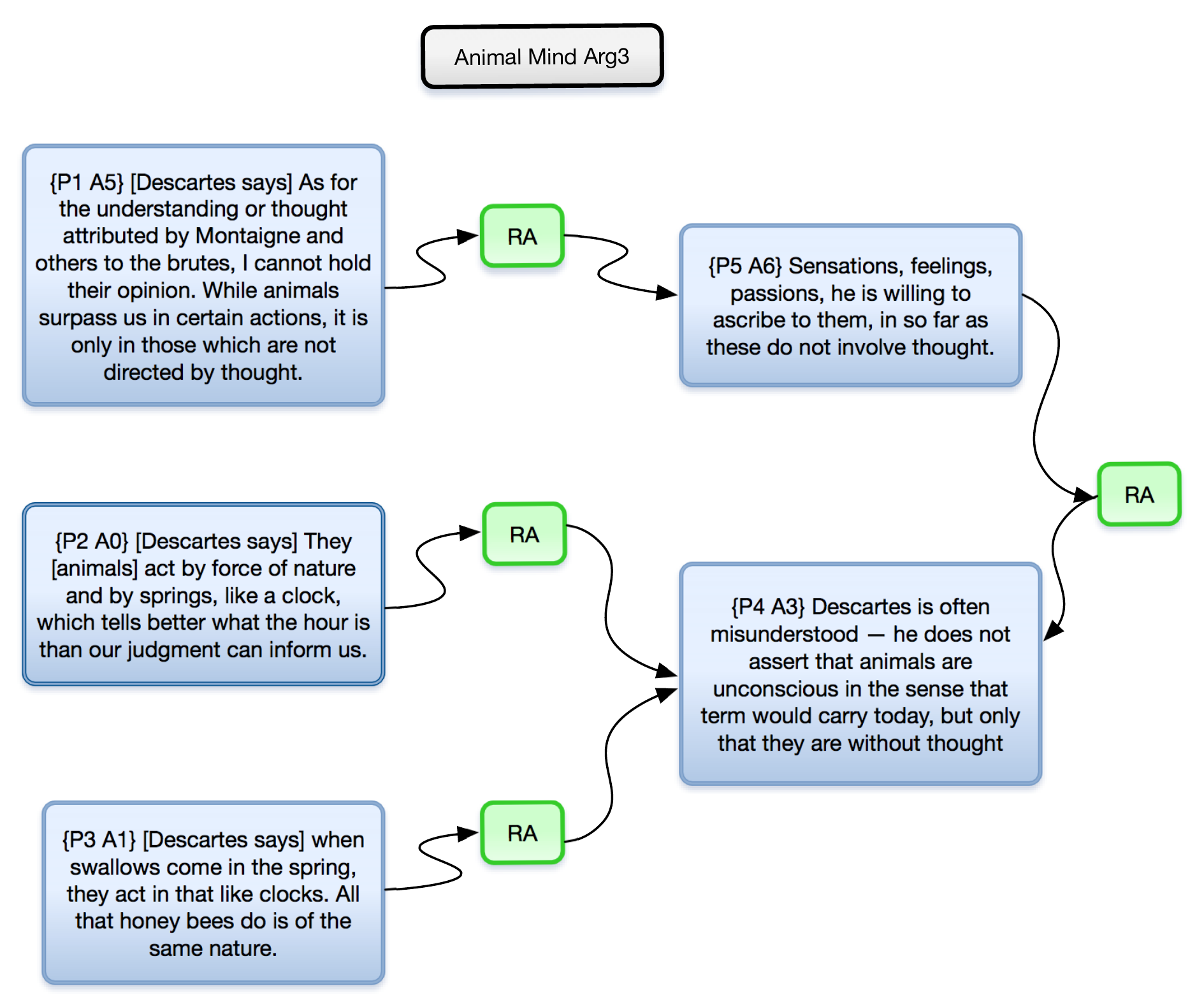}
\caption{An argument map derived from \emph{The Animal Mind}, represented in OVA+.}
\label{fig:argument}
\end{figure}

Using the $k=60$ model of the $HT6$ collection to identify sections of each book with highest proportion of Topic 1, we selected 108 pages from the six $HT6$ volumes for further analysis (Table~\ref{table:ht6-maps-pages}). From these we generated 43 argument maps using AIF annotated documents, providing a visual representation of the structure of each argument (e.g., Fig.~\ref{fig:argument}).

\begin{table*}
\begin{tabularx}{\textwidth}{|p{0.25\linewidth}|p{0.4\linewidth}c|c|c|}
\hline 
\textbf{Volume}  & \textbf{Pages} & \textbf{Total} & \textbf{Pass 1} & \textbf{Pass A}
\\ \hline
\emph{The Animal Mind} & 13-16, 16-21, 24-27, 28-31, 31-34, 58-64, 204-207, 288-294 & 40 & 9 & 15 \\
% uc2.ark+=13960=t5w66bs1h == 333 pages (.1201)
\emph{The Psychology of Ants} & Preface, 15-19, 31-34, 48-53, 99-103, 108-112, 206-209, 209-214 & 37 & 8 & 10 \\ %*
% uc2.ark:/13960/t6057f659 ==  200 pages(0.185)
\emph{The Principles of Heredity} & 374, 381, 382, 385, 386, 390, 394, 395 & 10  & & 8\\ % uc2.ark:/13960/t74t6gs0m ==379 pages (.0263)
\emph{General Biology} & 434-435, 436 & 3 & & 2 \\
% t0ht2h954 = 542 (.0055)
\emph{The Nature and Development of Animal Intelligence} & 16-18, 21-26, 30-32 & 12 & & 5\\ %*
% t05x26n0d = 307 (.0390)

\emph{Progress of Science} & 479-484 & 6 & & 3  \\ \hline %*
% t5p84550z = 536 (.0111)
\textbf{Overall Totals} & & 108 & 17 & 43 \\
\hline 

\end{tabularx}
\caption{Pages for which OVA+ argument maps were created, showing total number of pages analyzed and numbers of arguments identified on each of the passes described in the main text. }
%*Pages marked with a * were renumbered due to a software error.
\label{table:ht6-maps-pages}
\end{table*}

We performed two types of argument analysis: Pass 1 aimed to \emph{summarize} the arguments presented in each volume. Pass A aimed to \emph{sequence} the arguments presented in each volume. All argument maps can be found at \url{http://bit.ly/1bwJwF9}. A full description of the study, including analysis of the arguments can be found in \cite{McAlister2014}, and is summarized in \cite{Lawrence2014}.

As a proof of concept, these arguments show the utility of new techniques for faceted search enabling access from a library of over 300,000 books to volume-level analysis of a subset of 1,315 books all the way down to page-level analyses of 108 pages for the purpose of identifying, encoding, modeling, and visualizing arguments. These argument diagrams function as a type of close reading, common in the humanities, where this approach is related to a range of work in human-computer argument modeling \cite{ravenscroft2000,Yuan2011} and argument mapping {Kirschner2012}. This approach also draws on a rich tradition of philosophical literature (reviewed in \cite{Reed2007}).

The query sentence (\textsc{Q}) that we chose from Argument 15 of the Washburn book is shown below with the first half dozen results (and their similarity scores).\footnote{It is important to remember that LDA topic modeling is a “bag of words” approach; i.e., it uses only an unordered list of words in each document. It has no information about word order, punctuation, or other formatting in the text, and some of the most common words are not included. The full sentences are shown here only to aid the reader.}
\begin{enumerate}
\item[\textsc{Q:}] Every statement that another being possesses psychic qualities is a conclusion from analogy, not a certainty; it is a matter of faith. (1.0000)
\item If any consciousness accompanies it, then the nearest human analogy to such consciousness is to be found in organic sensations, and these, as has just been said, must necessarily be in the human mind wholly different in quality from anything to be found in an animal whose structure is as simple as the Amoeba's.	(0.8413)
\item Fancy, for example, one of us entering a room in the dark and groping about among the furniture. (0.8239)
\item This, of course, does not refer to the power to judge distance. (0.8235)
\item Again, a bodily structure entirely unlike our own must create a background of organic sensation which renders the whole mental life of an animal foreign and unfamiliar to us. (0.8224)
\item She disposes of the psychic learning by experience theory of Nagel by saying that the only experience upon which the animal could reject the filter paper must be experience that it is not good for food. (0.8198)
\item We speak, for example, of an ``angry'' wasp (0.7924)
\end{enumerate}

Sentence 1 is obviously related in meaning to the query sentence: the sentences overlap in some words, and directly express related ideas. But the relevance of the other examples is less direct. Sentence 6 provides a nice illustration of anthropomorphic attribution with no word overlap whatsoever. The inclusion of sentences 2 and 3 is, more puzzling. However, in the context of where these sentences appear in Washburn’s book, the relationship become plainer. Sentence 2 comes in the context of the discussion of what it might be like to be an amoeba. It is thus related to sentence 1, and it is used by Washburn to make the point that our experience in the dark, which still involves visual imagination and memories of what we touch, must be ``wholly different in quality'' (per sentence 1) from what an amoeba might experience. Sentence 3 occurs in a footnote on page 238, and it is worth quoting the footnote in full:
\begin{quote}
Porter observed that the distance at which spiders of the genera Argiope and Epeira could apparently see objects was increased six or eight times if the spider was previously disturbed by shaking her web (612). This, of course, does not refer to the power to \emph{judge} distance. [Italics in original.]
\end{quote}

Here, then, we see the author cautioning the reader not to jump to a high-level interpretation of the spider behavior. The spiders may perceive objects at various distances but they don’t judge it.  The term ‘judge’ here is philosophically interesting, as it suggests an influence of Immanuel Kant on framing the debate. While Kant's name does not appear in Washburn's book, the term ‘judgment’ is important to Kant’s theory of cognition, and fundamental to the cognitive divide he posits between humans and animals. We emphasize that this is just a speculative suggestion about Washburn’s influences, but it does show how the topic modeling process can bring certain interpretive possibilities to the fore, moving the digital humanities another step closer to the goal of generating new insight into human intellectual activity.

\begin{figure}
\includegraphics[width=7.0in]{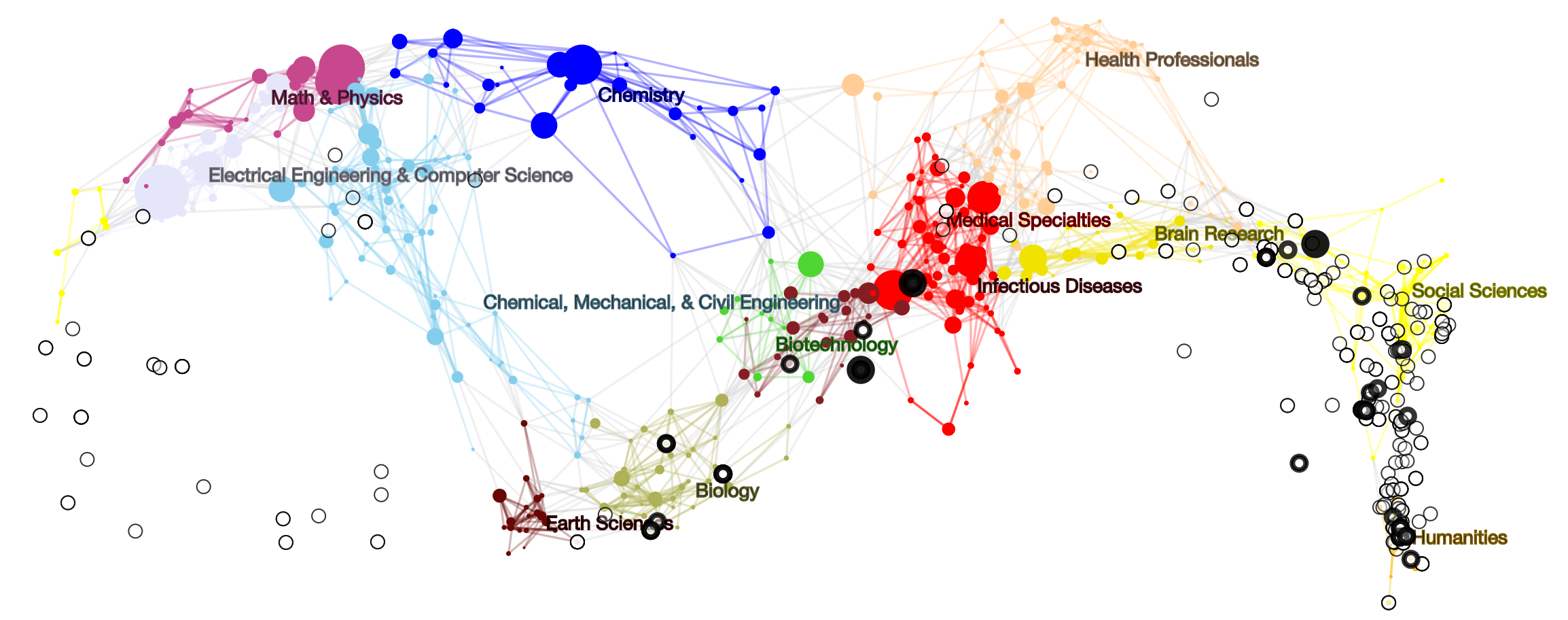}
\caption{UCSD Map of Science with overlay of HathiTrust search results shows topical coverage of humanities and life science data. The basemap of science shows each sub-discipline denoted by a circle colored or shaded according to the 13 core disciplines. Links indicate journal co-citations from the basemap. The 776 volumes of $HT1315$ with LCCN metadata are shown on the map as circles. Volumes also in $HT86$ are shown with thicker circles, and those in $HT6$ are shown in the thickest circles. An online, interactive version can be explored at \url{http://inpho.cogs.indiana.edu/scimap/scits}.%: Use scroll wheel or trackpad to zoom in and out, hover over a circle to display book title(s), click to see book record in the HT collection. 
}
\label{fig:crosswalk}
\end{figure}

Finally, we found Library of Congress classification records for 776 out of 1,315 books and we used the LCCO classification crosswalk that we constructed to locate these books on the UCSD Map of Science, as shown in Fig.~\ref{fig:crosswalk}.\footnote{In the interactive online version, nodes can be selected, showing which volumes are mapped and providing the title and links to various external sources of metadata.} The map confirms that the initial keyword-based selection from the HathiTrust retrieved books that are generally positioned below the “equator” of the map, with particular concentrations in the life sciences and humanities, as was to be expected. The map provides additional visual confirmation that the further selections via topic modeling to the $HT86$ and $HT6$ corpora managed to target books in appropriate areas of interest. 

Ultimately, the map overlay provides a grand overview and a potential guide to specific books that were topic modeled, although without further guidance from the topic models, the map does not fully meet the desired objective of linking a high-level overview to more detailed textual analysis.

\section*{General Discussion}
The notion of ``distant reading''~\cite{Moretti2013} has captured the imagination of many in the digital humanities. But the proper interpretation of large-scale quantitative models itself depends on having a feel for the texts, similar to Barbara McClintock's stress on having a ``feeling for the organism''~\cite{Keller1983} or Richard Feynman on the importance for nascent physicists of developing ``a `feel' for the subject'' beyond rote knowledge of the basic laws~\cite{Feynman1964}. The interpretation of data and models, whether in science or the humanities, is itself (as yet, and despite a few small successes in fields such as medical diagnosis) a task at which humans vastly outperform machines. For this reason, the digital humanities remain a fundamentally hermeneutic enterprise~\cite{Rockwell2016}, and one in which distant readings and close readings must be tightly linked if anything is to make sense.

In this paper we have motivated, introduced, and exemplified a multi-level computational process for connecting macro-analyses of massive amounts of documents to micro-level close reading and careful interpretation of specific passages within those documents. Thus we have demonstrated how existing computational methods can be combined in novel ways to go from a high-level representation of many documents to the discovery and analysis of specific arguments contained within documents. 

We have also shown how to zoom out to a macro-level overview of the search results. We presented a novel classification crosswalk between the Library of Congress Classification Outline (LCCO) and the UCSD Map of Science, which was constructed using only journal data, to extend the data to books. Because of the mismatch between the book data and the journal metadata, the crosswalk is not perfect, and the method of averaging locations places many books in uninterpretable regions of the map. Nevertheless, the visualization provides some useful information about the effectiveness of a simple keyword search in locating items of interest within a collection of hundreds of thousands of books. 

That our method succeeded in discovering texts relevant to a highly specific interdisciplinary inquiry shows its robustness to inconsistent and incomplete data. The HathiTrust Digital Library had OCR errors in 2.4\% of volumes as of May 2010 \cite{Conway2010}. While the quality of the HathiTrust has increased in the intervening years, it is still a pervasive issue in digital archives \cite{Kichuk2015}.

Multi-level topic modeling combined with appropriate measures of distance can efficiently locate materials that are germane to a specific research project, going from more than a thousand books, to fewer than a hundred using book-level topic models, and further narrowing this set down to a small number of pages within a handful of books using page-level topic models. The similarity measures used to span the word-topic and topic-document distributions are mediated by the topics in the model, and because every topic assigns a probability to every word in the corpus, this approach is highly adept at finding implicit relationships among the documents. Typical applications of topic modeling used elsewhere, such as graphing the rise and fall of topics through time, may show large-scale trends, but do not directly facilitate the interplay between distant reading and close reading that leads to deeper understanding. By connecting abstract, machine-discovered topics to specific arguments within the text, we have shown how topic modeling can bridge this gap.

Could the similar outcomes have been accomplished with alternative methods such as Latent Semantic Analysis (LSA; \cite{LandauerDumais1997}) or other Natural Language Processing (NLP) methods? We have investigated LSA as a discovery tool in the $HT1315$ and $HT86$ corpora. For users conditioned by online search engines, the word-centric search paradigm of LSA is more familiar than the topic-centric retrieval methods introduced here. Nonetheless, once users become aware of the potential for topic models, they provide more information about retrieved documents. For example, Washburn's \emph{The Animal Mind} is significant for its mixture of topics.\footnote{The topic models of the $HT1315$ collection can be explored using this volume as an entry point at the InPhO Topic Explorer page here: \url{http://bit.ly/2qQZCJm}} Indeed the most prominent topic in the book (Topic 42: light, eye, two, eyes, visual, fig, vision, red, distance, movement) points to the importance of perceptual systems in her discussion of anthropomorphic attributions, and it is noteworthy that her dismissal of amoeba consciousness is grounded in what she argues is the lack of crossmodal associations (between, e.g., touch and vision) that are part of the content of human perceptual experiences---viz. the sentence previously discussed concerning why our human experience in the dark must be ``wholly different in quality'' from what an amoeba might experience.

\section*{Conclusions}
Using a six-step process we progressively reduced 300,000 public domain volumes from the HathiTrust Digital Library to the 1,315 books in the $HT1315$ corpus, to the roughly 32,000 pages in the $HT86$ collection, to the over 17,000 sentences of the $HT6$ collection, to smaller set of the 108 pages selected for close reading and argument markup. This reduction allowed us to filter out discussions of anthropomorphic deities and zero in on key elements of late 19th and early 20th Century arguments about anthropomorphizing nonhuman organisms. 

The application to the history of comparative psychology described in our case study was successful in at least three ways. First, team members unfamiliar with the domain of comparative psychology were effectively guided towards highly relevant material in an unreadably large set of books. Second, team members were introduced to the work of Margaret Washburn, a pioneer of scientific comparative psychology, who wrote an important textbook of the early 20th Century~\cite{Washburn1908}---a book that went through four editions in as many decades, but has been largely forgotten since then (although see \cite{Washburn2010} for a recent tribute). Third, close reading of the arguments in the $HT6$ corpus revealed the surprising taxonomic range of these arguments, to include consideration even of consciousness in amoebae. This was surprising even to the domain experts on the team. This discovery raises new questions about the context for contemporary discussions of slime mold intelligence~\cite{Nakagaki2000, Reid2016} and opens up new avenues for research and analysis. By putting words into context, topic modeling enabled researchers to zero in on passages worthy of detailed analysis and further humanistic interpretation.  

It took Charles Darwin 23 years to read a number of books comparable in size to $HT1315$, as documented in his Reading Notebooks. At the unlikely rate of one book a day, it would take nearly four years to read this set of books in its entirety. Even allowing for the fact that one fifth of the volumes retrieved in our were course catalogs,  eliminating those would nonetheless leave a daunting, if not quite Olympian, reading task. As the majority of the volumes selected by keyword search were not directly relevant to the research project, the payoff made possible by more sophisticated computational analysis of the full texts was critical for the present task of finding arguments about anthropomorphizing animals. 

Although we do not claim that our methods provide the only possible approach, we are unaware of any other approach providing the kind of multilevel macro-to-micro approach needed to assist digital humanities researchers who wish to leverage computational methods against large data collections to support both distant reading and more traditional close reading analyses in the humanities. Nevertheless, these methods suffer from the problem already noted by Blei\cite{Blei2012} and others that objective methods for evaluating the quality of topics are lacking. We regard this as a function of the complexity of human semantic understanding---the holy grail of strong Artificial Intelligence. And while studies such as ours raise more questions than they answer, they point the way to even deeper understanding of large text corpora and more useful tools to scholars.

\section*{Availability of data and materials}

\paragraph*{Software} %The LoC-UCSD crosswalk is available on GitHub at \url{https://github.com/inpho/loc-ucsd}.

The InPhO Topic Explorer and \texttt{vsm} semantic modeling library are available on GitHub at \url{https://github.com/inpho/topic-explorer} and \url{https://github.com/inpho/vsm}, respectively. Both are published under the MIT License, a maximally-permissive open-source license.

The notebooks used for this analysis are available on GitHub at \url{https://github.com/inpho/digging}. All notebooks are released under a Creative Commons Attribution 4.0 International (CC BY 4.0) License.

%\section*{Competing interests}
%The authors declare no competing interests.

\section*{Acknowledgments}
This work was funded by the National Endowment for Humanities (NEH) Office of Digital Humanities (ODH) Digging Into Data Challenge (``Digging by Debating''; PIs Allen, B\"orner, Ravenscroft, Reed, and Bourget; award no. HJ-50092-12). The authors thank the Indiana University Cognitive Science Program for continued supplemental research funding, and especially for research fellowships for Jaimie Murdock and Robert Rose. We also thank the HathiTrust Research Center (HTRC) for their support of research activities and generous access to materials.

\section*{Authors' contributions}
CA and KB conceived the project initially and CA motivated the research for the history and philosophy of comparative psychology. CA, KB, AR, CR, and DB designed the workflow process. RR, DR and JO initially designed and programmed the \texttt{vsm} topic model implementation, and JM contributed to its further development. SM carried out the argument analysis using the OVA+ tool built by CR and JL, and a protocol devised by CA, DB, and AR. RL and JM constructed the LoC-UCSD crosswalk. JM built the interactive version of the UCSD map of science, with design help from RL and KB. JM, CA, RL, KB, and AR wrote the paper, borrowing in places from the project white paper and NEH final report written by SM.

\singlespacing

\bibliographystyle{unsrt}
\bibliography{new-citations}
\end{document}